\documentclass[conference]{IEEEtran}
\IEEEoverridecommandlockouts
\usepackage{cite}
\usepackage{amsmath,amssymb,amsfonts}
\usepackage{algorithmic}
\usepackage{graphicx}
\usepackage{textcomp}
\usepackage{xcolor}
\def\BibTeX{{\rm B\kern-.05em{\sc i\kern-.025em b}\kern-.08em
    T\kern-.1667em\lower.7ex\hbox{E}\kern-.125emX}}

\usepackage[T1]{fontenc}
\usepackage[utf8]{inputenc}
\usepackage{graphicx}
\graphicspath{ {./fig/} }

\usepackage[graphicx]{realboxes}
\usepackage{array}
\usepackage{multirow}
\usepackage{longtable}
\usepackage[hidelinks]{hyperref}
\usepackage{booktabs}
\usepackage{svg}
\usepackage{listings}
\usepackage{xcolor}
\usepackage{booktabs}
\usepackage{bbding}

\definecolor{codegreen}{rgb}{0,0.6,0}
\definecolor{codegray}{rgb}{0.5,0.5,0.5}
\definecolor{codepurple}{rgb}{0.58,0,0.82}
\definecolor{backcolour}{rgb}{0.95,0.96,0.96}
\lstdefinestyle{lststyle}{
    backgroundcolor=\color{backcolour},   
    commentstyle=\color{codegreen},
    keywordstyle=\color{magenta},
    numberstyle=\tiny\color{codegray},
    stringstyle=\color{codepurple},
    basicstyle=\ttfamily\footnotesize,
    breakatwhitespace=false,         
    breaklines=true,                 
    captionpos=b,                    
    keepspaces=true,                 
    numbers=left,                    
    numbersep=5pt,                  
    showspaces=false,                
    showstringspaces=false,
    showtabs=false,                  
    tabsize=2
}
\begin{document}

\title{Towards Interoperability of Open and Permissionless Blockchains: A Cross-Chain Query Language
\thanks{This work is partially supported by the Swiss National Science Foundation project Domain-Specific Conceptual Modeling for Distributed Ledger Technologies [196889].}
}

\author{\IEEEauthorblockN{Felix Härer}
\IEEEauthorblockA{\textit{Digitalization and Information Systems Group} \\
\textit{University of Fribourg}\\
Fribourg, Switzerland \\
E-Mail: felix.haerer@unifr.ch}
}

\maketitle

\begin{abstract}
The rise of open and permissionless blockchains has introduced novel platforms for applications based on distributed data storage. At the application and business levels, long-established query languages such as SQL provide interoperability that can be complemented by blockchain-based data storage today, enabling permissionless and verifiable data storage along with decentralized execution across tens of thousands of nodes. However, when accessing one or more blockchains, interoperability is not provided today, posing challenges such as inhomogeneous data access in addition to different features and trade-offs, e.g. in data and distribution, scalability, and security. Towards interoperability in data access among the increasing number of blockchain platforms, this paper introduces a cross-chain query language for data access across blockchains. Similar to SQL, the language abstracts from implementation based on a data model compatible with the largest open and permissionless blockchains (OPB) today. The language, data model, and processing architecture are demonstrated and evaluated with an implemented prototype, aiming to contribute to the discussion on blockchain interoperability among OPB.
\end{abstract}


\section{Introduction}
\label{sec:introduction}

In August 2022, at least 22 well-known openly accessible blockchains exist with a significant number of active participants, processing their transactions over internet protocols\footnote{\url{https://messari.io/screener/active-adresses-15D9AE99}}. These systems are, in their nature, open platforms for distributed data and applications, allowing applications of businesses or individuals to transact through a shared ledger without centralized coordination based on algorithmic consensus in the network~\cite{belottiVademecumBlockchainTechnologies2019}. Open systems with these components - blockchain data, network, consensus protocol - are verifiable in their operation and data by all participants, enabling novel decentralized applications such as programmable money or contracts \cite{antonopoulosMasteringEthereumBuilding2019}. In contrast to distributed systems of past decades, hundreds to thousands of nodes establish open and permissionless platforms. Estimations based on the connections of participating nodes count about 15000 nodes operating Bitcoin\footnote{\url{https://bitnodes.io/}}, 6000 operating Ethereum\footnote{\url{https://ethernodes.org/}}, and 3000 operating Cardano\footnote{\url{https://adastat.net/pools/}}; without accounting for additional nodes not visible due to their configuration or placement behind routers and firewalls. With the growing adoption, the increasing number of open and permissionless blockchains, and the vast amounts of openly available data, future relevance of these platforms for verifiable data storage and execution is assumed as a premise of this paper. Applications accessing the platforms include payments and currency, e-commerce, timestamping, and the attestation of data and links on the web \cite{narayananBitcoinAcademicPedigree2017,weberProgrammableMoneyNextgeneration2021,harerDecentralizedAttestationDistribution2022}. 

\emph{Problem Statement.} Software accessing data across open and permissionless blockchains (OPB) today faces challenges due to interoperability:

\begin{enumerate}
    \item Inhomogeneous access to data due to various OPB implementations.
    \item Different OPB data models and features exist.
    \item Different OPB trade-offs exist, notably regarding scalability, security, and decentralization.
\end{enumerate}

\emph{Research Objective and Contribution.} The objective of this research is to address the three challenges towards higher interoperability between OPB. In the discussion on this topic, the paper contributes a cross-chain query language by specifying a common data model, a standardized syntax, and a processing architecture. For answering query statements written by users or applications, data is read from multiple blockchain nodes, integrated in the common data model, and processed according to the statements. Given the state-of-the-art of conceptual models and query languages suggested before, such as \cite{oliveConceptualSchemaEthereum2020} and \cite{bragagnoloEthereumQueryLanguage2018}, the language design is derived abstract from implementation for several of the largest OPB today. The proof-of-concept implementation demonstrates feasibility and the potential for software to utilize OPB as part of their architecture.

\emph{Application Example.}
Several e-commerce websites might take part in commonly operated loyalty programs that issue reward points for customer purchases. Among other industries, these programs can be found for airlines working together\footnote{E.g. \url{https://www.miles-and-more.com/ch/en.html}}. Given a cross-chain query language, business-level applications of different airlines could re-use and standardize data access, the underlying blockchain could be changed, and multiple blockchains of several loyalty programs could be used from individual applications.

The remainder of this paper is structured as follows. Section \ref{rel} introduces foundations and related work. Section \ref{ql} discusses OPB with their properties for the derivation of a data model. Subsequently, the data model, a grammar of the language syntax, and a processing architecture are derived. This approach is evaluated in Section \ref{ev} with a prototype implementation utilizing multiple OPB. Section \ref{co} concludes.

\section{Foundations and Related Work}
\label{rel}

Blockchain foundations are introduced first, followed by open and permissioned blockchains and interoperability.

\subsection{Blockchains}

Following the initial publication of the Bitcoin whitepaper and software in 2008 and 2009, respectively \cite{nakamotoBitcoinPeertoPeerElectronic2008,nakamotoBitcoinSoftwarebasedOnline2009}, the term \emph{blockchain} has been introduced as a generalization of its technical architecture. The main components of (1.) a data structure of backward-linked blocks, (2.) a peer-to-peer network for data distribution, and (3.) a consensus protocol allow for novel properties. Notably, coordinating and validating all operations without centralized control, open access to all data and operations, and permissionless access where data and operations are not restricted to specific participants~\cite{garayBitcoinBackboneProtocol2015,antonopoulosMasteringEthereumBuilding2019}. Ethereum, and other blockchains following it, extended the capabilities with smart contracts as quasi Turing-complete programs~\cite{woodEthereumSecureDecentralised2022,buterinEthereumNextGenerationSmart2014}. In addition to payments and money, smart contracts enable e-commerce, sales contracts, timestamping, and attestations, among other applications~\cite{ladleifUnifyingModelLegal2019a,narayananBitcoinAcademicPedigree2017,harerDecentralizedAttestationDistribution2022}. 

\subsection{Open and Permissionless Blockchains}
\label{opb}

Growing development and adoption originating from Bitcoin and Ethereum have resulted in OPB with various properties. Table \ref{tab:opb} lists five well-known OPB in order of their public number of network nodes, characterized by the properties of their data structure, network, and consensus protocols, as well as features related to smart contracts.

\begin{table*}[]
\caption{Properties of well-known open and permissionless blockchains.}
\scalebox{1.00}{
\begin{minipage}{\textwidth}
\centering
\begin{tabular}{lllll}
\toprule
Blockchain &
  Data Structure &
  \begin{tabular}[c]{@{}l@{}}Network\end{tabular} &
  Consensus Protocol &
  \begin{tabular}[c]{@{}l@{}}Smart Contract\\Features\end{tabular} \\ \midrule
{[}1{]} Bitcoin\footnote{[18], \url{https://bitnodes.io/}} &
  \begin{tabular}[c]{@{}l@{}}Blocks, UTXO \\ data model\end{tabular} &
  \begin{tabular}[c]{@{}l@{}}Bitcoin, approx. 
  \\ 15000 nodes\end{tabular} &
  \begin{tabular}[c]{@{}l@{}}Nakamoto Consensus,\\ 
  Proof-of-Work\end{tabular} &
  \begin{tabular}[c]{@{}l@{}}Stack-based script\\ execution, monetary\\ transactions\end{tabular} \\
{[}2{]} Ethereum\footnote{[29], \url{https://ethereum.org/en/developers/docs/}, \url{https://ethernodes.org/}} &
  \begin{tabular}[c]{@{}l@{}}Blocks, account \\ state storage in tree \\ data structures\end{tabular} &
  \begin{tabular}[c]{@{}l@{}}Ethereum Mainnet, \\ approx.
  \\ 6000 nodes\end{tabular} &
  \begin{tabular}[c]{@{}l@{}}Ethash, memory-hard \\ 
  Proof-of-Work\end{tabular} &
  \begin{tabular}[c]{@{}l@{}}Ethereum Virtual \\ Machine, general-\\ purpose programs\end{tabular} \\
{[}3{]} Cardano\footnote{[15], \url{https://adastat.net/pools/}} &
  \begin{tabular}[c]{@{}l@{}}Blocks, extended \\ UTXO model\end{tabular} &
  \begin{tabular}[c]{@{}l@{}}Cardano, \\ approx.
  \\ 3000 nodes\end{tabular} &
  \begin{tabular}[c]{@{}l@{}}Ouroboros, \\ 
  Proof-of-Stake\end{tabular} &
  \begin{tabular}[c]{@{}l@{}}General-purpose \\ programs, \\ functional\end{tabular} \\
{[}4{]} Solana\footnote{[30], \url{https://docs.solana.com}, \url{https://solanabeach.io/validators/}} &
  \begin{tabular}[c]{@{}l@{}}Block and graph \\ data structures over \\ different time spans\end{tabular} &
  \begin{tabular}[c]{@{}l@{}}Solana Mainnet Beta, \\ approx.
  \\ 1600 nodes\end{tabular} &
  \begin{tabular}[c]{@{}l@{}}Graph-based (proof-\\ -of-history), Proof-of-\\ Stake\end{tabular} &
  \begin{tabular}[c]{@{}l@{}}General-purpose\\ programs\end{tabular} \\
{[}5{]} Avalanche\footnote{[23], \url{https://stats.avax.network/dashboard/network-status/}\\} &
  \begin{tabular}[c]{@{}l@{}}Block and graph\\ data structures over \\ different networks\end{tabular} &
  \begin{tabular}[c]{@{}l@{}}Platform/Exchange/\\ Contract (P/X/C) \\ chain, approx.
  \\ 1300 nodes\end{tabular} &
  \begin{tabular}[c]{@{}l@{}}Avalanche (P Chain)\\ Snowman (X/C Chain), \\ Proof-of-Stake\end{tabular} &
  \begin{tabular}[c]{@{}l@{}}Ethereum Virtual \\ Machine (C Chain), \\ general-purpose \\ programs\end{tabular}
\end{tabular}
\end{minipage}
}
\label{tab:opb}
\vspace{-5mm}
\end{table*}

\emph{Data Structures.}
The initial design of backward-linked blocks in Bitcoin is combined in most other OPB with additional trees or graphs. In addition to transaction data from blocks, further queries must be carried out for non-transactional data or older data that has been pruned. For example, separate tree structures are present for state storage in Ethereum, where balances and smart contract variables can be retrieved \cite{oliveConceptualSchemaEthereum2020}.

\emph{Networks.}
Major OPB networks consist of approximately 1300 to 15000 nodes. Due to algorithmic operation and validation, higher node counts increase security, e.g. related to 51\% attacks and selfish mining \cite{shrivasDisruptiveBlockchainSecurity2020,saadExploringAttackSurface2020} frequently observed in small Proof-of-Work systems such as \emph{Bitcoin Gold}~\cite{saadExploringAttackSurface2020}.

\emph{Consensus Protocols.} In the protocols initially created between 2008 and 2022, a shift away from Proof-of-Work to Proof-of-Stake can be observed, introducing several trade-offs. While established blockchains such as Bitcoin and Ethereum have been emphasizing security and decentralization over the past years, efforts to improve efficiency and scalability are made in Cardano \cite{kiayiasOuroborosProvablySecure2016a}, Avalanche \cite{rocketScalableProbabilisticLeaderless2020}, and Solana \cite{yakovenkoSolanaNewArchitecture2018}. The tendency is reflected in work on novel consensus protocols by the three blockchains, mostly based on Proof-of-Stake~\cite{giladAlgorandScalingByzantine2017,ethereumProofofstakePoS2022} with higher efficiency and advantages to environmental impact. Avalanche and especially Solana emphasize scalability. In Solana, however, temporary protocol failures can be observed frequently, resulting in non-availability \cite{haywardSolanaBlamesDenial2021}.

\emph{Smart Contract Features.} For data queries and software applications, smart contract features are required. In this area, Bitcoin possesses a limited scripting language used for programmable monetary transactions and the scalable \emph{lightning} overlay network~\cite{antonopoulosMasteringLightningNetwork2021}. The introduction of general-purpose programming in Ethereum and others introduces greater features and complexity. At present, most implementations are written and compiled for the Ethereum Virtual Machine, present in Ethereum and Avalanche. Cardano and Solana possess their own architectures with support for general-purpose programs.

\subsection{Blockchain Interoperability}
\label{interop}

Interoperability is broadly recognized for transactions across blockchains in cross-chain swaps and similar concepts found in practice in so-called \emph{bridges}. In addition, standardization efforts related to inhomogeneous data are beginning, not limited to query languages.

\emph{Cross-Chain Swaps.} Swaps are commonly initiated through a protocol on an initial blockchain, where tokens or arbitrary data are locked to prevent further transfer in the beginning. A counterparty transaction is issued on a second blockchain to the initiator of the cross-chain swap, i.e., often another party pays for the tokens with another asset on the second chain. This transaction includes a cryptographic proof with a secret that unlocks the tokens on the initial chain. Finally, the counterparty withdraws tokens from the initial chain. Various protocols on this basis and variants exist~\cite{pillaiCrosschainInteroperabilityBlockchainbased2020,shadabCrosschainTransactions2020}. In atomic cross-chain swaps \cite{herlihyAtomicCrossChainSwaps2018a,zakharyAtomicCommitmentBlockchains2020}, atomicity is provided for all transfers involved in a cross-chain swap. Implementations in bridges may, in practice, exhibit differing properties and assurances, not necessarily providing atomicity or other guarantees for the completion of the exchange. Bridges exist mainly for cryptocurrency exchange, e.g. Anyswap\footnote{\url{https://anyswap.exchange/}} and Connext\footnote{\url{https://bridge.connext.network/}} allow for cross-chain swaps between Ethereum, Avalanche, and others. Cross-chain swaps and bridges are not standardized and do not provide homogeneous access or queries.

\emph{Inhomogeneous Data.} Standardization efforts address inhomogeneous data with few prior works related to inhomogeneous access. For Ethereum, \cite{oliveConceptualSchemaEthereum2020} discusses a conceptual schema derived from the main data structures of the blockchain. \cite{bragagnoloEthereumQueryLanguage2018} propose a query language for the content of blocks and transactions. The language design is based on SQL in its syntax and supports concepts such as projection and selection within Ethereum. For data analysis, a framework and implementation based on Scala has been proposed~\cite{bartolettiGeneralFrameworkBlockchain2017}, where SQL or NoSQL is used with aggregation functions and similar analysis methods. The analysis approach \cite{camozziMultidimensionalAnalysisBlockchain2022} describes a data warehouse and ETL process for analyzing Ethereum data using standard SQL with a multi-dimensional data model for queries of dimension attributes and data aggregation support. This work and similar works might connect to multiple blockchains, however, they do not provide homogeneous data access, queries, or simultaneous access to data of multiple blockchains. Other works based on SQL include \cite{liivExplorationStructuredQuery2021}, using multiple blockchains for populating a standard MySQL database with the third-party service Google BigQuery. The use of third-party services as data sources presents another problem often observed in prior work, where validation of blockchain data is not possible or severely limited. Further approaches include public connectors between blockchains, blockchains that integrate with others, and hybrid approaches~\cite{belchiorSurveyBlockchainInteroperability2021a}.

\emph{Limitations of Prior Work.} In conclusion, present solutions are limited regarding (L1.) homogeneous data access, (L2.) standardized queries, (L3.) simultaneous access to multiple blockchains, or (L4.) blockchain data validation. Currently, the focus is on cross-chain swaps and siloed data analysis rather than data integration. The proposed query language addresses these limitations by suggesting a common data model (L1.), a standardized syntax (L2.), and a processing architecture (L3.) supporting local nodes (L4.) of multiple blockchains.

\section{Query Language}
\label{ql}

The two following subsections describe (A.) the data model and (B.) the syntax and processing architecture of the language. Query statements are processed according to the architecture in subsection (B.), resulting in instances of data model classes using data provided by APIs of local blockchain nodes. 

\subsection{Data Model}

\begin{figure*}[ht]
    \centering
    \includegraphics[width=1.00\linewidth]{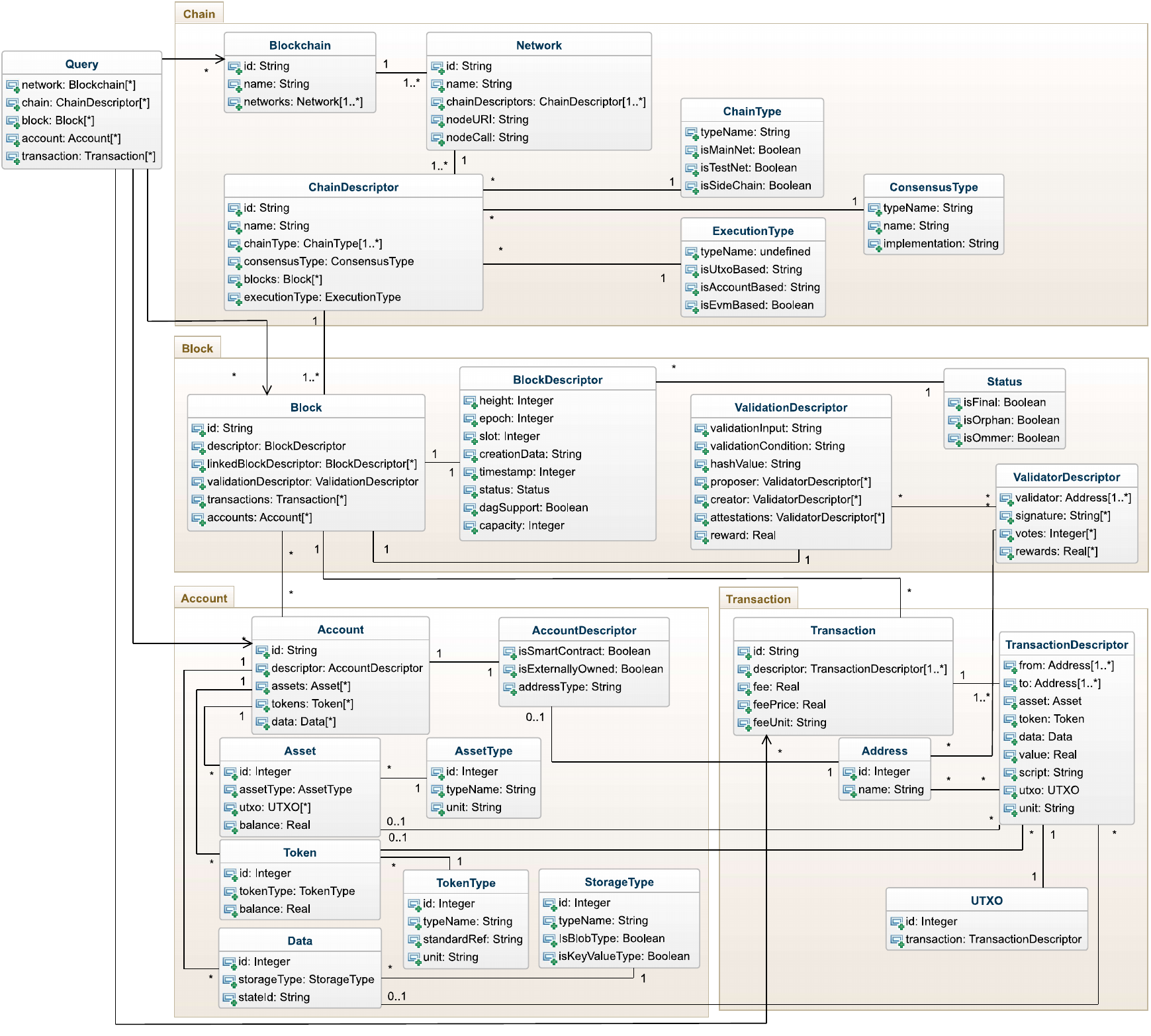}
    \caption{Data model of the cross-chain query language as UML class diagram.}
    \label{fig:data-model}
    \vspace{-2mm}
\end{figure*}

The language design is based on a data model integrating the main data structures and attributes of the OPB introduced in Section \ref{opb}. Based on prior work and existing tools discussed in Section \ref{interop}, classes and attributes of the five OPB have been identified, generalized, and integrated in a common data model. Figure~\ref{fig:data-model} lays out the complete data model as UML class diagram. Table~\ref{tab:opb-support} lists the main model classes grouped into four packages for representing the chain, block, account, and transaction concepts of the OPB. For formulating queries, the syntax is introduced in subsection~\ref{lang-syntax-proc-architecture}. Statements are written in terms of the classes and attributes by specifying the source data with class and attribute names of the data model.

\begin{table*}[ht]
\centering
\caption{Data model classes supporting the concepts of the OPB.}
\scalebox{1.00}{
\begin{tabular}{@{}llll@{}}
\toprule
Chain classes & Block classes & Account classes  & Transaction classes \\ \midrule

Bitcoin chain & Blocks of transactions & - & Values (Bitcoin), UTXO  \\

Ethereum chain & Blocks of transactions & Data storage, balances & Values (Ether), \\
                   &                        &   & data incl. tokens  \\

Cardano chain & Epochs, slots with     & Addresses, data & Values, assets, \\
                     & blocks for transactions &                 & data, UTXO\\

Solana chain & Epochs, slots for transactions & Data storage & Data incl. tokens\\

Avalanche P/X/C chain & P/X: transaction DAG & P/X: - & X: values, assets, UTXO\\
  
            & C: blocks of transactions & C: Data storage, balances & C: Data incl. tokens
  \\ \bottomrule \\
\end{tabular}
}
\label{tab:opb-support}
\vspace{-1mm}
\end{table*}

In the table and data model, the concepts of the OPB are represented by the following classes. The chain classes represent one main network and blockchain for Bitcoin, Ethereum, Cardano, and Solana, by the classes Chain, Network, and ChainDescriptor of the data model. Additional test networks with their separate blockchains, e.g. Ropsten and Görli in Ethereum, are represented by Network and ChainDescriptor. In Avalanche, the Network class encompasses one primary network, the first of potentially many \emph{subnets}, with separate ChainDescriptor instances for the three P/X/C blockchains.

The Block and BlockDescriptor classes represent blocks with separate classes for the status of the block, the block validation through the consensus protocol, and the validators involved. Conceptually, blocks are described by an ID in the form of a hash value in all blockchains, with metadata such as timestamp and a height denoting the block number under the assumption no changes occur to non-final blocks. For example, in Bitcoin, multiple blocks could be found as the successor to a given block; however, only one block will be included in the chain while the others will be discarded with \emph{orphan} status. In contrast, the same case is handled in Ethereum by keeping one block in the main chain and keeping the other blocks at the same level with \emph{ommer} status. Blocks are not explicitly finalized in Proof-of-Work chains, allowing for the declaration of \emph{orphan} or \emph{ommer} status for blocks found in parallel to prior blocks of the chain. However, the probability of existing blocks being replaced in this way decreases over time since multiple successive parallel blocks with greater cumulative work are required. An explicit finalization of blocks, preventing the occurrence of multiple successors, can be found in more recent Proof-of-Stake blockchains such as Solana.

Regarding data structure, blocks are linked to one or more existing blocks of the linkDescriptor attribute of the Block class, establishing either backward-linked blocks or a graph, such as a directed acyclic graph (DAG) in the Avalanche C chain. Blocks either contain transactions directly or are grouped into time-based slots and epochs for validation purposes with Proof-of-Stake. When appending a block, validation is carried out for each block or slot, requiring the participation of validators. In the ValidationDescriptor class, the creator of a block for Bitcoin and Ethereum provides validation for a linked block by the hashValue attribute. For the other Proof-of-Stake blockchains, proposers are recorded in the corresponding attributes with attestations, referencing the class ValidatorDescriptor. Each instance references any number of appointed validators attesting the correctness of the block by means of their vote and signature. In this way, the concept of multiple groups of validators performing attestations is represented. When storing the transaction of a DAG, one or more transactions in a block can be linked to one or more transactions from a preceding block, indicated by the linkedBlockedDescriptor attribute in Block and the dagSupport attribute in BlockDescriptor that are set \emph{true} in this case.

Accounts are a concept present in Ethereum, Solana, and Avalanche. Blocks contain accounts for the storage of assets, tokens, or data used for smart contracts. Notably, data might be used for the representation of assets or tokens directly, such as in Solana. Each account is described by an ID with the concept of an address being present in all blockchains. In an account, the storage of assets or tokens can refer to custom assets, such as in Cardano, or tokens represented by data in the general case. For tokens, token standards such as ERC-20 in Ethereum are represented by Token class attributes. Storing data uses binary large objects or key-value stores, utilized in hash-based mapping data structures.

Transaction concepts in Bitcoin and Cardano differ due to the lack of accounts in these blockchains. For this reason, transactions contain a reference to unspent transaction outputs (UTXOs) of previous transactions. In this model, a UTXO is included together with the transferred value and a script describing locking conditions or containing data. While the inclusion of data is implicit in Bitcoin, Cardano explicitly supports data in transactions and its storage linked to an address for smart contract functionality. In the case of Ethereum, Solana, and the Avalanche C chain, transactions are stored for the transfer of values, data, assets, or tokens between accounts. In the Avalanche X chain, the transfer of native assets is supported through the UTXO concept. In the data model, the attributes of Transaction and TransactionDescriptor allow for transfers between addresses by utilization of the attributes corresponding to the aforementioned concepts.



\subsection{Language Syntax and Processing Architecture}
\label{lang-syntax-proc-architecture}

The language syntax is based on established concepts of data query languages, in particular the Structured Query Language (SQL). On the one hand, the syntax of SQL and similar languages allow the representation of queries in a formalized way through relational algebra. On the other hand, queries and their articulation are accessible to domain experts without detailed knowledge of the underlying concepts. The SQL syntax is centered around the \emph{SELECT-FROM-WHERE} block (SFW block). Based on English-language commands, the \emph{SELECT} clause will perform a projection in the underlying relational model, semantically corresponding to columns, followed by the source of the relations in the \emph{FROM} clause and the selection of tuples using conditions in the \emph{WHERE} clause. In the relational model, set operations, and notably the Cartesian product, are the basis for all queries. For a cross-chain data language, these concepts are applied as follows. 

\medskip
\emph{Query Requirements.} The syntax of query statements consists of query (Q), source (S), and filter (F) clauses:

\begin{itemize}
    \item[Q] Query attributes are any attributes of the data model classes. Each attribute must be specified together with its class, determining one column of the query result for each source.
    \item[S] Sources specify blockchains and networks, optionally together with blocks, transactions, and accounts with assets, tokens, and data. In terms of the data model, each source must be specified by the attribute values of the identifying attributes of the Chain, Network, and ChainDescriptor classes. Optionally, an additional class, attribute, and attribute value of an identifying attribute from the classes Block, Transaction, Account, Asset, Token, or Data can be specified. 
    \item[F] Filters are optional conditions filtering the query results by query attributes and sources. Each filter must be specified by a filter function for comparisons with two inputs using query attributes. On the result values from the query attributes, specified filters are applied in sequence.
\end{itemize}

\begin{figure*}[ht]
    \centering
    \includegraphics[width=0.75\linewidth]{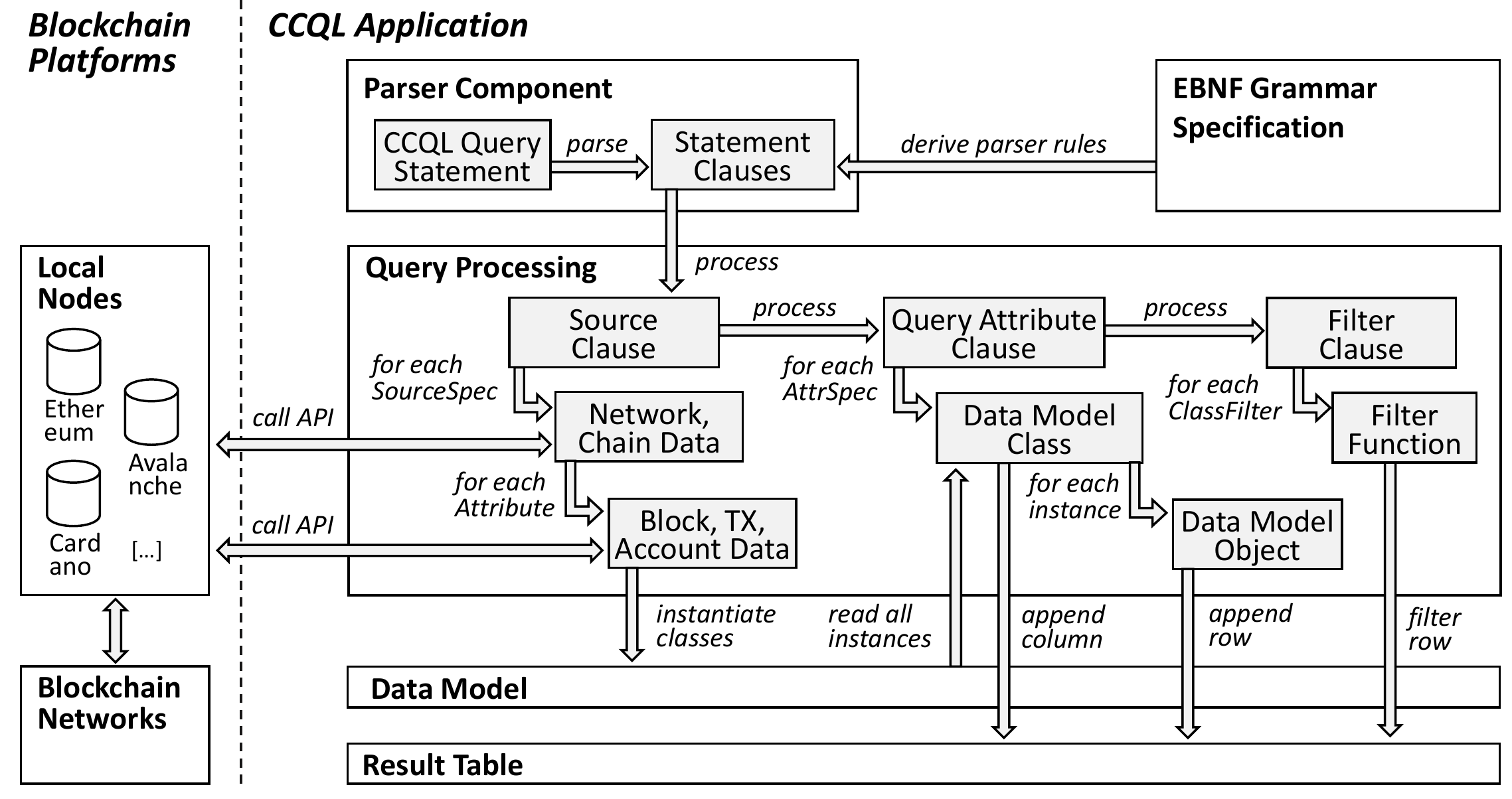}
    \caption{Architecture for processing queries within an application.}
    \label{fig:arch-query-proc}
    \vspace{-3mm}
\end{figure*}

\medskip
\emph{Grammar.} According to these requirements, Listing~\ref{lst:grammar1} partially shows the syntax definition\footnote{Complete grammar:  \url{https://github.com/fhaer/CCQL/tree/main/grammar}}. In the grammar excerpt, the query statement syntax is described using the W3C variant of the Extended Backus Naur Form (EBNF)~\cite{w3cBlindfoldGrammars2001}. The query, source, and filter clauses specify projections, source data, and selections, respectively. In each clause, the specification of multiple values entails processing multiple result sets, in the case of SourceSpec for triggering the retrieval of data from multiple blockchains. Additionally, sources specify a chain, network, and chain descriptor with an optional block, transaction, or account according to the data model and requirements. Accounts storing assets, tokens, or data are accessed by the respective classes shown in the data model. The source and filter specifications are further detailed with the full EBNF grammar specification in the implementation. 

\begin{lstlisting}[caption={EBNF excerpt. Attr: Attribute, Spec: Specification, Val: Value, Desc: Descriptor, I: Instance, Net: Network, Tx: Transaction, Acc: Account.},label={lst:grammar1}]
QueryStatement ::= 
  QueryAttrClause 
  SourceClause
  FilterClause? ";"
QueryAttrClause ::= 
  'Q ' AttrSpec ( ', ' AttrSpec )*
SourceClause ::=
  'S ' SourceSpec ( ', ' SourceSpec )*
FilterClause ::=
  'F ' FilterSpec ( ', ' FilterSpec )*
AttrSpec ::=
  CCQLClass '.' AttrName
SourceSpec ::=
  BlockchainI ':' NetI ':' ChainDescI 
  (':' ( BlockI | TxI | AccI ) )?
FilterSpec ::= 
  CCQLClass '.' AttrName ComparisonFunction IValue
\end{lstlisting}

\medskip
\emph{Processing Architecture.} The processing of queries within the architecture is described in Figure~\ref{fig:arch-query-proc}. In an application connected to local nodes of blockchain platforms, query statements are issued to the parser component. After building the clauses, the query processing component first iterates the source clause, issuing for each SourceSpec and for each attribute API calls to the blockchain nodes. According to the classes and attributes, instances are created in the data model. Second, the query clause iterates each AttrSpec by selecting the corresponding classes from the model and producing columns in the result table. For the objects of the classes, rows are appended. Third, the filter clause applies each filter function. Finally, the result table contains the query result.

\section{Evaluation of Feasibility}
\label{ev}

The aim of this section is a demonstration of feasibility for the query language, its data model, and the processing architecture. For this purpose, an implementation compatible to the introduced OBP has been developed, consisting of a formal language grammar and a prototype application\footnote{Available at \url{https://github.com/fhaer/CCQL/tree/main}}. The grammar is realized with the Eclipse Modeling Framework and Xtext\footnote{\url{https://www.eclipse.org/Xtext}} to establish an external domain-specific language (DSL) based on it. In principle, the grammar is implementation independent and might be re-used in further applications. The language is implemented in a prototype command-line application together with the data model according to the proposed architecture, involving node access to the selected OPB for query execution. The prototype is written in Python 3.9 and utilizes the web3.py library for OPB access\footnote{\url{https://web3py.readthedocs.io/en/stable/}}. 

\subsection{Software Setup}

Setting up the application involved the following blockchain nodes with a configuration that fully validates all blocks:

\begin{itemize}
    \item Bitcoin node: Bitcoin Core, version 22.0\footnote{\url{https://bitcoin.org/de/download}}. Initial data synchronization completed after 4 days.
    \item Ethereum node: go ethereum (geth\footnote{\url{https://geth.ethereum.org/downloads/}}), version 1.10.510 with tracing and indexing of all transactions, and pruning of ancient block data. Initial data synchronization completed after approximately 12 weeks.
    \item Cardano node: Cardano node, version 1.34\footnote{\url{https://github.com/input-output-hk/cardano-node}}. Initial data synchronization  completed after approximately 2 days.
    \item Avalanche node: AvalancheGo, version 1.77\footnote{\url{https://github.com/ava-labs/avalanchego/releases}}. Initial data synchronization completed after approximately 4 days.
\end{itemize}

\begin{figure*}[ht]
    \centering
    \includegraphics[width=1.01\linewidth]{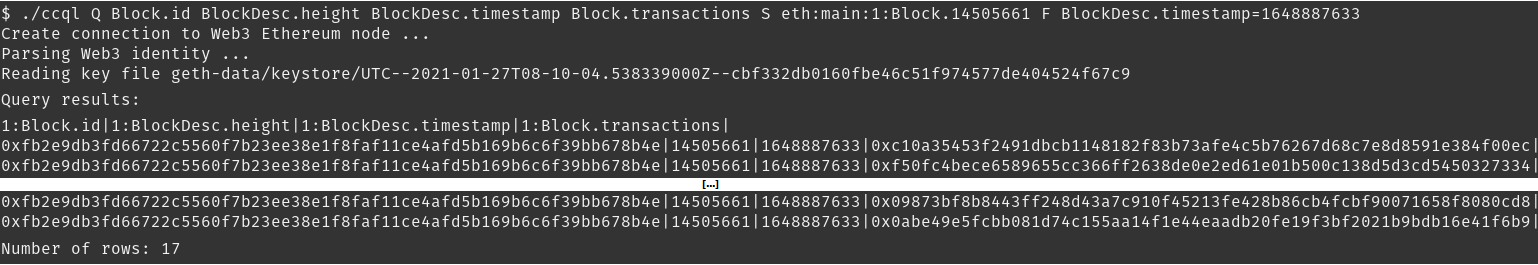}
    \caption{Query example 1.}
    \label{fig:ev1}
\end{figure*}

\begin{figure*}[ht]
    \centering
    \includegraphics[width=1.01\linewidth]{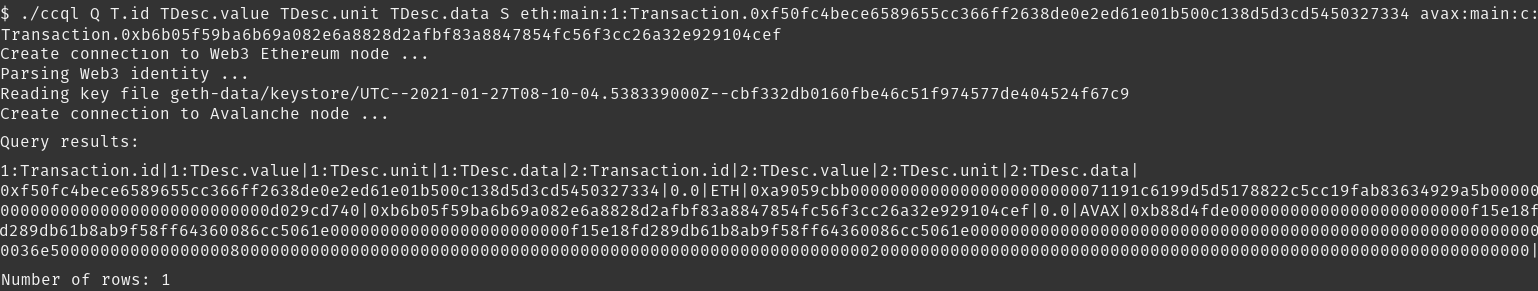}
    \caption{Query example 2.}
    \label{fig:ev2}
\end{figure*}

The synchronizations were made on a PC with a AMD 3700X CPU, 32 GB RAM, and Samsung 980 Pro NVMe SSD, behind a 1 Gbit/s fiber internet connection. After the synchronizations, nodes and data were transferred to a typical end-user device, a laptop with AMD 5700U CPU, 16 GB RAM, and SK Hynix BC711 NVMe SSD. Both machines were running Ubuntu 21.10.

In order to demonstrate feasibility, query statements were evaluated with the prototype, as discussed in the following section. Each statement was executed on the laptop device, with local data retrieval from the blockchain node software according to the processing architecture. Given the local and fully-validating configuration of the node software, query results can be produced without network access. Query performance, therefore, does not depend on network latency and is limited only by local CPU and IO performance. 

\subsection{Prototype Operation}

The architecture of the prototype is represented in Figure~\ref{fig:arch-query-proc} and generally explained in Section~\ref{lang-syntax-proc-architecture}, \emph{Processing Architecture}. In the following, two example queries are discussed.

The common task of locating transactions in a block is shown in Figure~\ref{fig:ev1}. The query attributes specify Block and BlockDescriptor (BlockDesc) classes with attributes of the block ID, Height, Timestamp, and transactions. Ethereum, its main network, chain 1, and block follow as source clause. Finally, a filter with a timestamp is applied. In the query results, data model classes and attributes such as Block.id and Block.height can be seen with their corresponding values, e.g. \texttt{0xfb2e\small{[...]}} and \texttt{14505661}, respectively.

With similar classes and attributes, the second example in Figure~\ref{fig:ev2} shows a query common in a cross-chain swap scenario, retrieving transaction data from Ethereum and Avalanche. The query results show the attributes prefixed by the number of the source with instance-level data from the data model with corresponding values, with blocks and transaction identifiers as hexadecimal hash values. 




\subsection{Discussion}

The prototype implements homogeneous data access to OPB, e.g. retrieving asset and data transfers from multiple blockchains. According to the grammar, data access is standardized and allows for statements involving one or more blockchains. For utilizing blockchain properties in a meaningful way, running blockchain nodes locally is required, involving substantial time and cost for initial synchronizations. The data model architecture follows a data integration approach, where data can be stored according to today's major OPB by populating relevant classes. In contrast, an approach of multiple individual data models alone would not address the problem at hand. General limitations of the prototype in its current stage are limited support for advanced concepts of the OPB, e.g. the calculation of transaction fees involving further utility tokens is outside the model. In terms of functionality, the queries with filters are limited in the prototype, only allowing for inner joins and equality comparisons. 

\section{Conclusion}
\label{co}

In this paper, a cross-chain query language grammar, data model, and processing architecture are introduced for homogeneous data access across multiple blockchains. The approach supports homogeneous data access, standardization of queries, addressing of multiple blockchains in individual queries, and local validation of blockchain data. Prior research only covered these aspects partially. The feasibility of the approach could be positively evaluated with a prototype, despite functional limitations present in the implementation. In principle, software applications can utilize such an integrated approach for storing data on open and permissionless blockchains, independent of their implementation and openly available. Future research will further evaluate and extend the language design, data model, and architecture toward execution capabilities.

\bibliographystyle{IEEEtranDOI}
\bibliography{references.bib}{}

\end{document}